\documentclass[12pt]{article}
\usepackage{epsfig}

\def\beq{\begin{equation}}
\def\eeq#1{\label{#1}\end{equation}}
\def\eeqn{\end{equation}}

\def\beqa{\begin{eqnarray}}
\def\eeqa#1{\label{#1}\end{eqnarray}}
\def\eeqan{\end{eqnarray}}

\let\bar=\overbar

\def\etal{{\it et al.}}

\def\Dslash{\not{\hbox{\kern-4pt $D$}}}
\def\dslash{\not{\hbox{\kern-2pt $\del$}}}

\def\msb{{\bar{\ssstyle M \kern -1pt S}}}

\def\Title#1{\begin{center} {\Large {\bf #1} } \end{center}}
\newenvironment{Abstract}{\begin{quotation}  }{\end{quotation}}

\def\support{\footnote{Speaker on behalf of the Belle Collaboration}}

\newcommand{\dtagkx}{D_{\rm tag}KX_{\rm frag}}
\newcommand{\ds}{D_{s}}

\newcommand{\dsst}{D^{\ast}_{s}}
\newcommand{\xfrag}{X_{\rm frag}}
\newcommand{\dssttodsg}{D^{\ast}_{s}\to\ds\gamma}

\newcommand{\dtag}{D_{\rm tag}}

\newcommand{\dz}{D^0}
\newcommand{\dc}{D^+}
\newcommand{\lc}{\Lambda_c^+}
\newcommand{\dsmunu}{\ds^+\to\mu^+\nu_{\mu}}

\newcommand{\br}{{\cal B}}
\newcommand{\fds}{f_{\ds}}
\newcommand{\fb}{fb$^{-1}$}
\newcommand{\ellnu}{\ell\nu_{\ell}}

\newcommand{\munu}{\mu\nu_{\mu}}
\newcommand{\taunu}{\tau\nu_{\tau}}
\newcommand{\taumunu}{\tau(\mu)\nu_{\tau}}

\newcommand{\mmiss}{M_{\rm miss}}

\newcommand{\eecl}{E_{\rm ECL}}

\begin{document}

\begin{center}
Proceedings of CKM 2012, the 7th International Workshop on the CKM Unitarity Triangle,\\
University of Cincinnati, USA,\\  
28 September - 2 October 2012
\end{center}

\bigskip\bigskip

\Title{\boldmath Leptonic $D_s$ decays at $B$-factories}

\bigskip\bigskip

\begin{raggedright}  

{\it An\v ze Zupanc\index{Zupanc, A.}\support\\
Karlsruher Institut f\"ur Technologie\\
Institut f\"ur Experimentelle Kernphysik\\
76131 Karlsruhe, Germany}\\
\bigskip\bigskip
\end{raggedright}

\begin{Abstract}
We review recent measurements of leptonic $D_s$-meson decays performed by Belle and BaBar collaborations. 
Described measurements enable experimental extraction of the $D_s$-meson decay constant which can be 
compared with lattice QCD calculations. 
\end{Abstract}

\section{Introduction}
The leptonic decays of mesons provide access to experimentally clean measurements of
the meson decay constants or the relevant Cabibbo-Kobayashi-Maskawa
matrix elements. In the Standard Model (SM) the branching fraction for a leptonic decay of 
a charged pseudoscalar meson, such as $D^+_s$, is given by \cite{PDG,Rosner:2012bb}:
\begin{equation}
 \br(D_{s}^+\to \ell^+\nu_{\ell})=\frac{\tau_{\ds}M_{D_{s}}}{8\pi}f_{D_{s}}^2G_F^2|V_{cs}|^2m_{\ell}^2\left(1-\frac{m_{\ell}^2}{M_{D_{s}}^2} \right)^2,
 \label{eq:brleptonic_sm}
\end{equation}
where $M_{D_{s}}$ is the $D_{s}$ mass, $\tau_{\ds}$ is its lifetime, $m_{\ell}$ is the lepton mass, $V_{cs}$ is the 
Cabibbo-Kobayashi-Maskawa (CKM) matrix element between the $\ds$ constituent quarks $c$ and $s$, and $G_F$ is the Fermi coupling constant. 
The parameter $f_{D_{s}}$ is the decay constant, and is related to the wave-function overlap of the quark and anti-quark. 
The magnitude of the relevant CKM matrix element, $|V_{cs}|$, can be obtained from the very well measured 
$|V_{ud}|=0.97425(22)$ and $|V_{cb}|=0.04$ from an average of exclusive and inclusive semileptonic B decay results
as discussed in Ref. \cite{vcb} by using the following relation, $|V_{cs}|=|V_{ud}|-\frac{1}{2}|V_{cb}|^2$. Measurements of leptonic 
branching fraction of a pseudoscalar meson thus provide a clean probe of the decay constant which can than be compared with precise
lattice QCD calculations~\cite{Na:2013ti}. 

\section{Absolute branching fraction measurement}
The methods of absolute branching fraction measurement of $\ds^-\to\ell^-\overline{\nu}{}_{\ell}$ decays used recently by Belle~\cite{Zupanc:2012cd} and
before by the BaBar \cite{Sanchez} are similar. Both collaborations study $e^+e^- \to c\bar{c}$ events which contain $\ds^-$ mesons 
produced through the following reactions:
\begin{equation}
 e^+e^-\to c\bar{c}\to \dtagkx\ds^{\ast -},~\ds^{\ast -}\to\ds^-\gamma. 
 \label{eq:signal_events_type}
\end{equation}
In these events one of the two charm quarks hadronizes into a $\ds^{\ast -}$ meson while the other quark
hadronizes into a charm hadron denoted as $\dtag$ (tagging charm hadron). The above events are reconstructed fully in two steps: in the first step
$\ds$ mesons are reconstructed inclusively while in the second step $\ds\to\ellnu$ decays are reconstructed within the inclusive sample.
The tagging charm hadron is reconstructed as $\dz$, $\dc$, $\lc$\footnote{In events where $\lc$ is reconstructed as tagging charm 
hadron additional $\overline{p}$ is reconstructed in order to conserve the total baryon number.} in 18 (15) hadronic decay modes by Belle 
(BaBar). In addition $D^{\ast +}$ or $D^{\ast 0}$ are reconstructed in order to clean up the event. 
The strangeness of the event is conserved by requiring additional kaon, denoted as $K$, which can be either $K^+$ or $K^0_S$. 
Since $B$-factories collected data at energies well above 
$D{}^{(\ast)}_{\rm tag} K D_s^{\ast}$ threshold additional particles can be produced in the process of hadronization. These particles are 
denoted as $\xfrag$ and can be: even number of kaons and or any number of pions or photons. 
Both Belle and BaBar reconstruct $\xfrag$ modes with up to three pions in order to keep background at reasonable level. $\ds^-$ mesons are required to be produced in a
$\ds^{\ast -}\to\ds^-\gamma$ decays which provide powerful kinematic constraint  ($\dsst$ mass, or mass difference between $\dsst$ and $\ds$) 
that improves the resolution of the missing mass (defined below) and suppresses the combinatorial background.

\begin{figure}[t]
 \centering
 \includegraphics[width=0.54\textwidth]{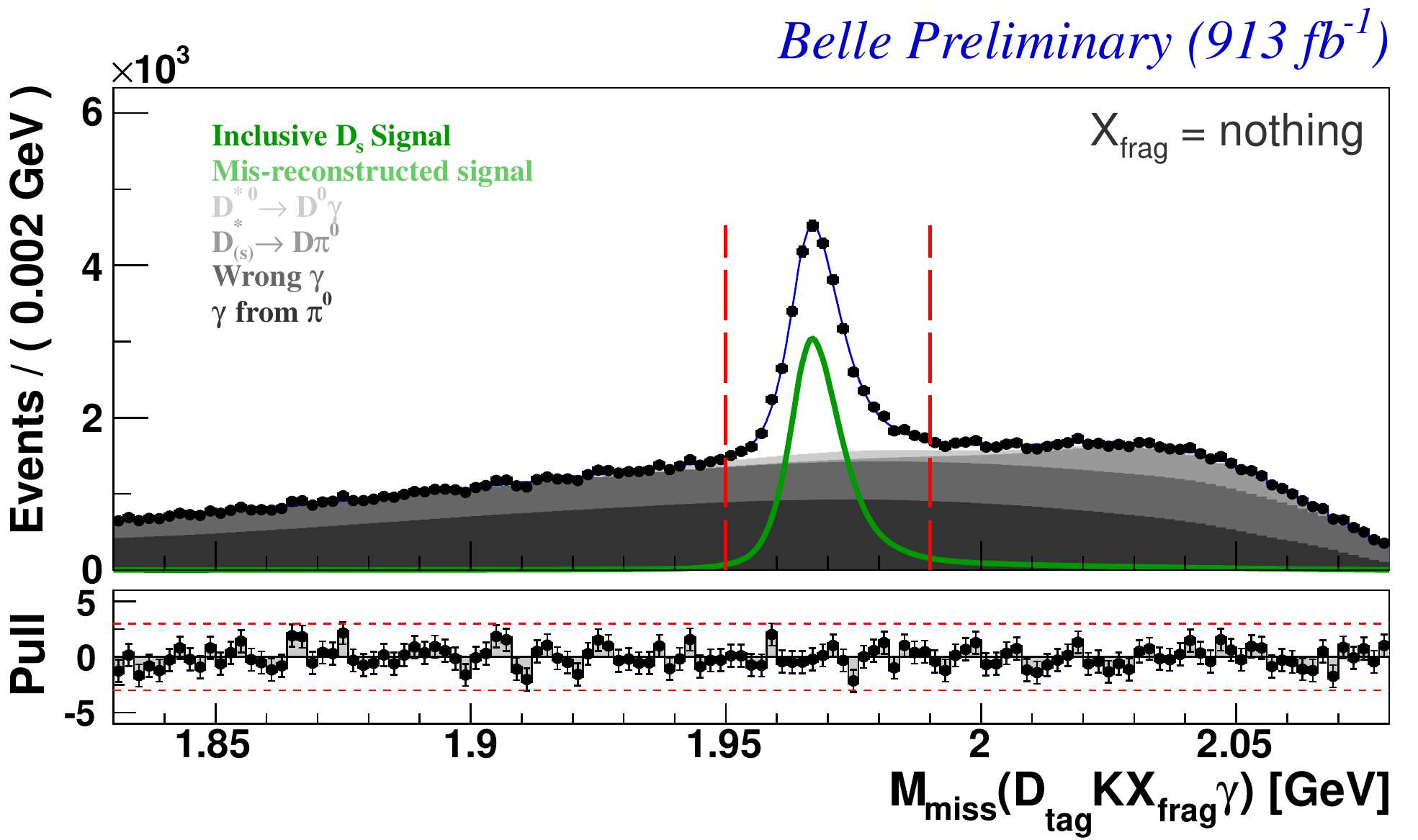}
 \includegraphics[width=0.45\textwidth]{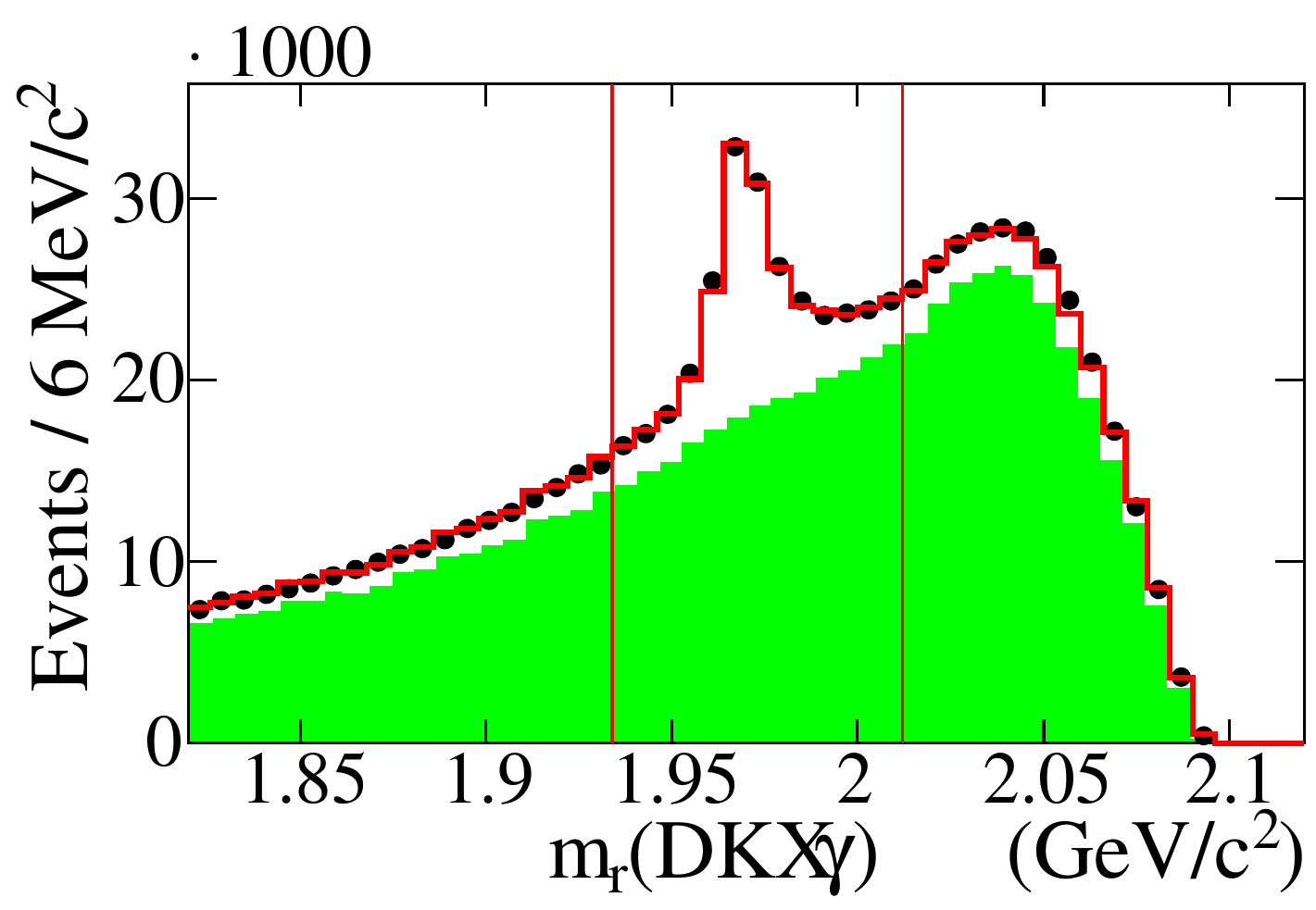}
 \caption{The $\mmiss(\dtagkx\gamma)=m_r(DKX\gamma)$ distributions for the cleanest $\xfrag$ mode (left) from Belle and all $\xfrag$ modes combined (right) from BaBar.
 The two vertical lines indicate the signal region used in the $\ellnu$ selections.}
 \label{fig:dsincl}
\end{figure}

In the first step of the measurement no requirements are placed on the daughters of the signal $\ds^-$ meson
in order to obtain a fully inclusive sample of $\ds^-$ events which is used for normalization
in the calculation of the branching fractions. The number of inclusively reconstructed $\ds$ mesons is
extracted from the distribution of events in the missing mass, $\mmiss(\dtagkx\gamma)$, recoiling against the $\dtagkx\gamma$ system:
\begin{equation}
\mmiss(\dtagkx\gamma)  =  \sqrt{p_{\rm miss}(\dtagkx\gamma)^2},
\label{eq:massds} 
\end{equation}
where $p_{\rm miss}$ is the missing momentum in the event:
\begin{equation}
p_{\rm miss}(\dtagkx\gamma)  =  p_{e^+} + p_{e^-} - p_{\dtag} - p_{K} - p_{\xfrag} - p_{\gamma}.\\
\label{eq:pmiss}
\end{equation}
Here, $p_{e^+}$ and $p_{e^-}$ are the momenta of the colliding positron and electron beams, respectively, and the $p_{\dtag}$, $p_{K}$, 
$p_{\xfrag}$, and $p_{\gamma}$ are the measured momenta of the reconstructed $\dtag$, kaon, fragmentation system and the photon from
$\dssttodsg$ decay, respectively. Correctly reconstructed events given in the Eq. \ref{eq:signal_events_type} produce a peak in the 
$\mmiss(\dtagkx\gamma)$ at nominal $\ds$ meson mass as shown in Fig. \ref{fig:dsincl}. Belle finds $94400\pm 1900$ correctly reconstructed
inclusive $\ds$ candidates in a data sample corresponding to 913 \fb, while BaBar finds $108900\pm2400$ events\footnote{
Note that Belle quotes number of correctly reconstructed candidates while BaBar number of events. It is subtle but important difference that
reader should be aware of.} containing $\ds$ meson in a data
sample corresponding to 521 \fb.

In the second step Belle and BaBar search for the purely leptonic $\ds^+\to\mu^+\nu_{\mu}$ and $\ds^+\to\tau^+\nu_{\tau}$ decays within the inclusive $\ds^+$ 
sample by requiring that there be exactly one additional charged track identified as an electron, muon or charged pion present in the rest of the event. 
In case of $\ds^+\to\tau^+\nu_{\tau}$ decays the electron, muon or charged pion track identifies the subsequent $\tau^+$ decay to 
$e^+\nu_e\overline{\nu}{}_{\tau}$, $\mu^+\nu_{\mu}\overline{\nu}{}{\tau}$ or $\pi^+\overline{\nu}{}_{\tau}$. 

\begin{figure}[tb]
 \centering
 \includegraphics[width=0.59\textwidth]{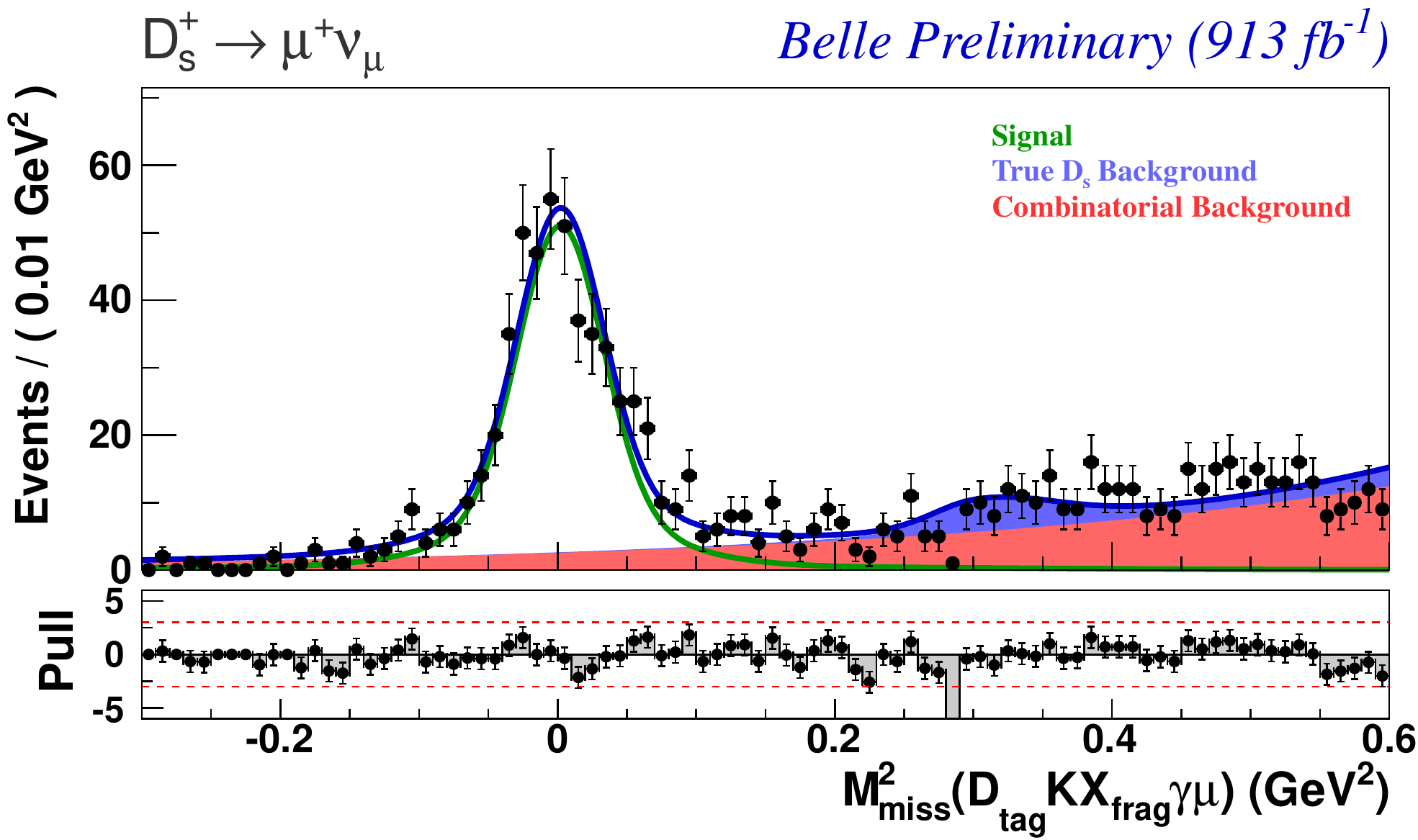}
 \includegraphics[width=0.4\textwidth]{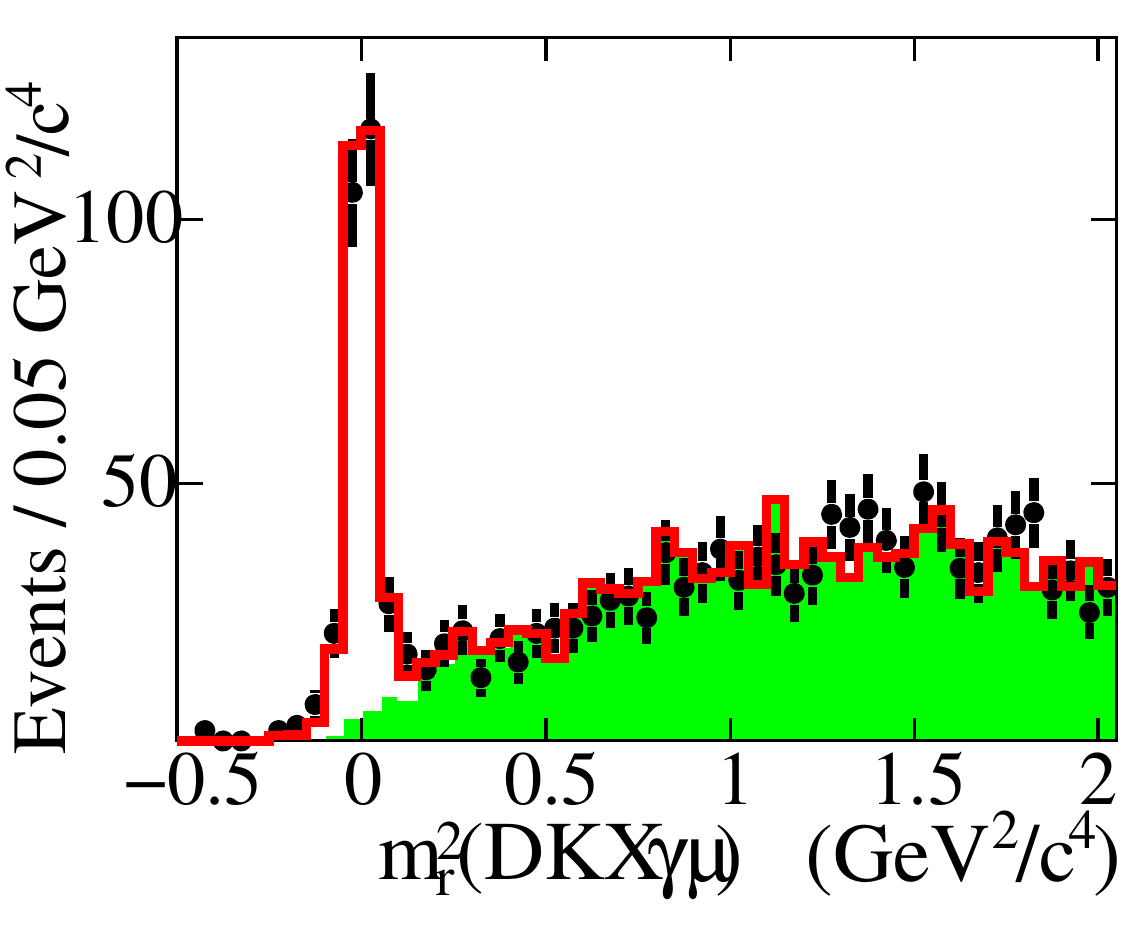}
 \caption{The $\mmiss^2(\dtagkx\gamma \mu)=m_r^2(DKX\gamma\mu)$ distributions for $\dsmunu$ candidates
 within the inclusive $\ds$ sample.}
 \label{fig:dsmunu}
\end{figure}

The $\dsmunu$ decays are identified as a peak at zero in the missing mass squared distribution, 
$\mmiss^2(\dtagkx\gamma \mu) = p_{\rm miss}^2(\dtagkx\gamma \mu)$ shown in Fig. \ref{fig:dsmunu}.

\begin{figure}[tb]
 \centering
 \includegraphics[width=0.57\textwidth]{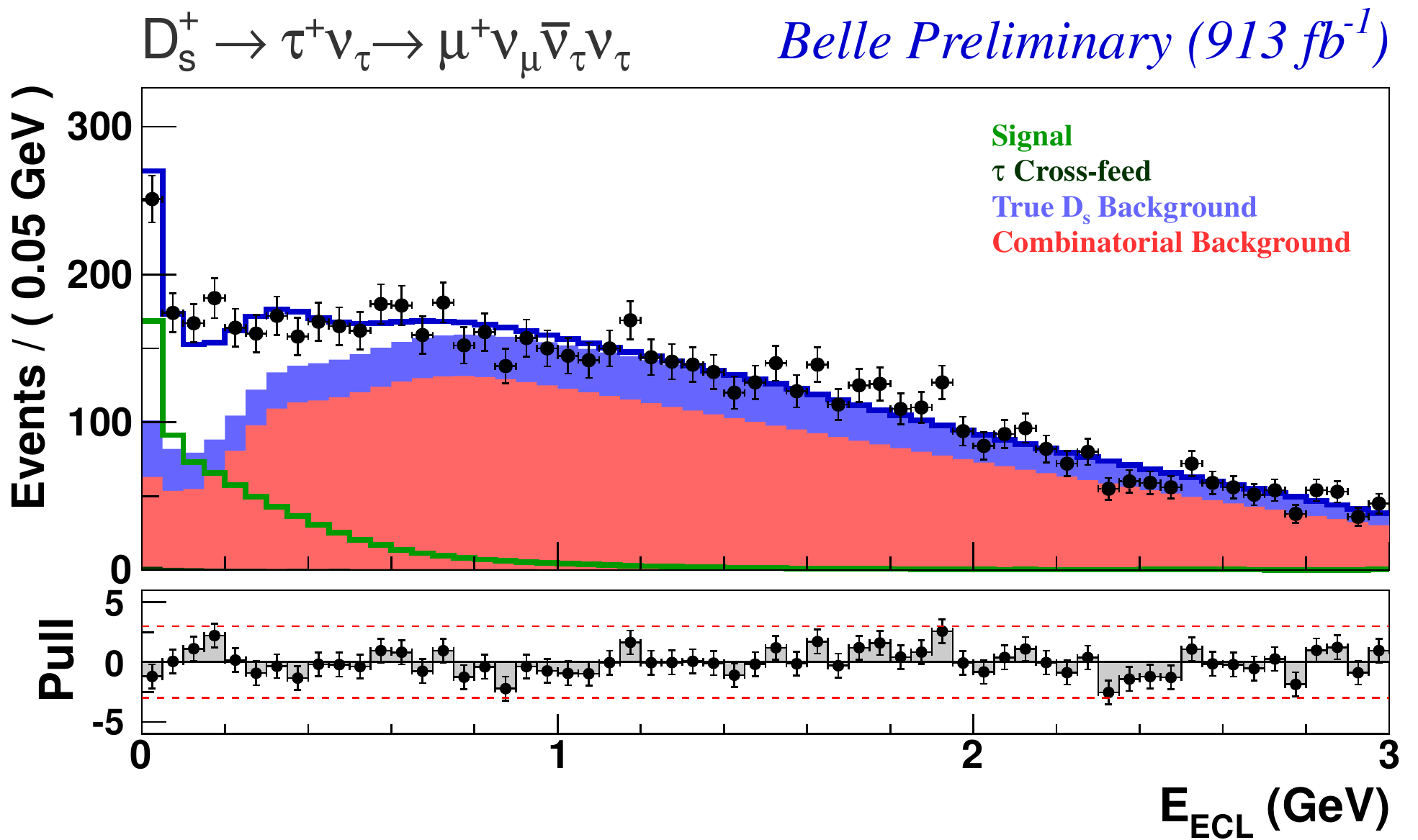}
 \includegraphics[width=0.42\textwidth]{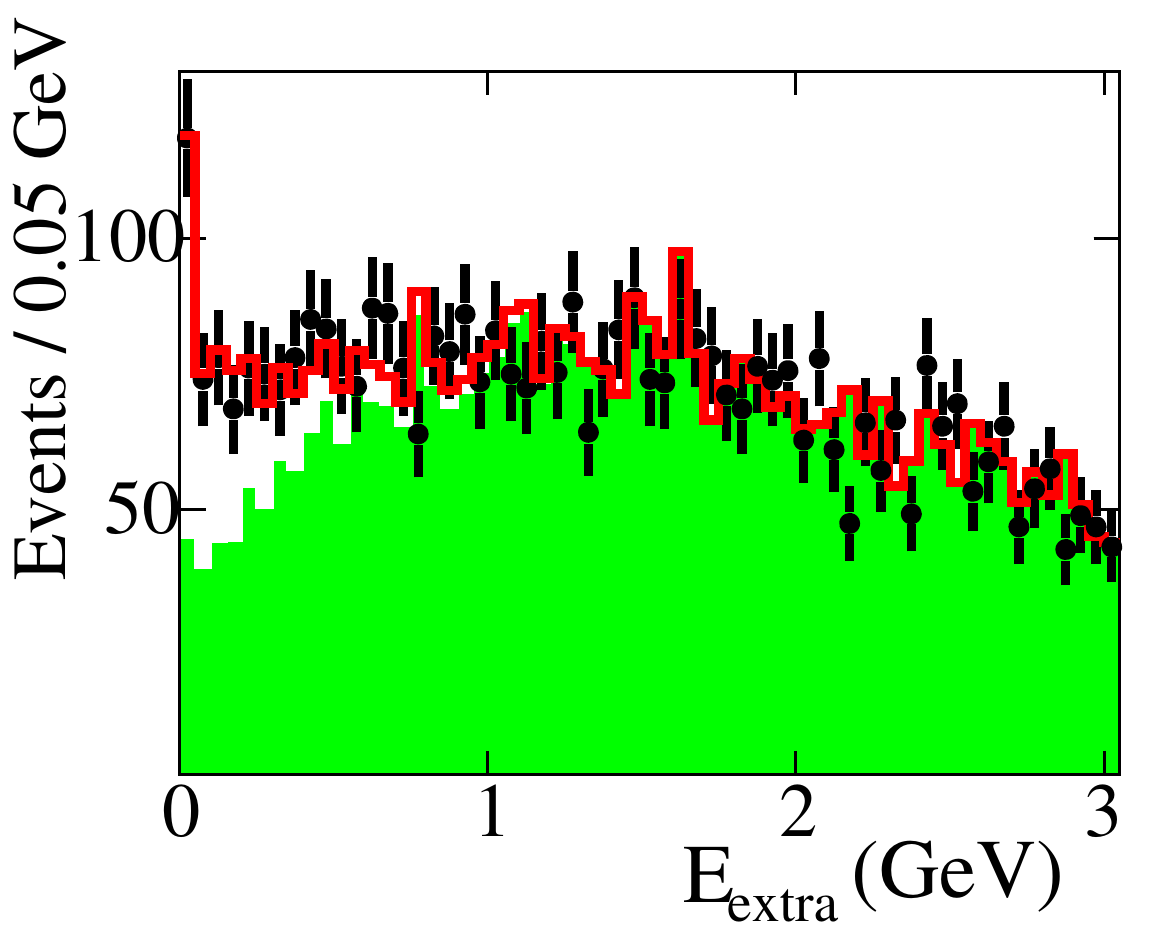}
 \caption{The $\eecl=E_{\rm extra}$ distribution for $\ds\to\taumunu$ candidates from Belle (left) and BaBar (right).}
 \label{fig:dstaunu}
\end{figure}

Due to multiple neutrinos in the final state the $\ds\to\taunu$ decays don't peak in the $\mmiss^2$ distribution. Instead, Belle and BaBar use extra
neutral energy in the calorimeter\footnote{The $E_{\rm ECL}$ ($E_{\rm extra}$) at Belle (BaBar) is defined as a sum over all energy deposits in the calorimeter 
with individual energy greater than 50 (30) MeV and which are not associated to the tracks or neutrals used in inclusive reconstruction of 
$\ds$ candidates nor the $\ds\to\taunu$ decays.}, $\eecl$, to extract the signal yields of $\ds\to\taunu$ decays. These are expected to peak
towards zero in $\eecl$, while the backgrounds extend over a wide range as shown in Fig. \ref{fig:dstaunu} for $\ds\to\taunu$ candidates
when $\tau$ lepton is reconstructed in its leptonic decay to a muon.

Table \ref{tab:results} summarizes the signal yields and measured absolute branching fractions of leptonic $\ds$-meson decays at Belle and BaBar.
The latter are found to be consistent within uncertainties.
\begin{table}[t]\footnotesize
 \centering
 \begin{tabular}{l|cc|cc}\hline\hline
  & \multicolumn{2}{|c|}{Belle} & \multicolumn{2}{|c}{BaBar}\\
  $\ds^+$ Decay Mode	& Signal Yield				& $\br$ [\%] 				& Signal Yield				& $\br$ [\%] \\
  \hline\hline
  $\munu$		& $\phantom{0}489\pm26\phantom{0}$	& $0.528\pm0.028\pm0.019$		& $\phantom{0}275\pm17\phantom{0}$	& $0.602\pm0.038\pm0.034$\\\hline
  $\taunu$ ($e$ mode)	& $\phantom{0}952\pm59\phantom{0}$	& $5.37\pm0.33{}^{+0.35}_{-0.30}$	& $\phantom{0}489\pm26\phantom{0}$	& $5.07\pm0.52\pm0.68$\\
  $\taunu$ ($\mu$ mode)	& $\phantom{0}758\pm48\phantom{0}$	& $5.88\pm0.37{}^{+0.34}_{-0.58}$	& $\phantom{0}489\pm26\phantom{0}$	& $4.91\pm0.47\pm0.54$\\
  $\taunu$ ($\pi$ mode)	& $\phantom{0}496\pm35\phantom{0}$	& $5.96\pm0.42{}^{+0.45}_{-0.39}$	& 	& \\\hline
  $\taunu$ (Combined)   & 					& $5.70\pm0.21{}^{+0.31}_{-0.30}$	& 					& $5.00\pm0.35\pm0.49$\\\hline\hline
 \end{tabular}
 \caption{Signal yields and measured branching fractions for $\ds^+\to\ell^+\nu_{\ell}$ decays by Belle and BaBar. 
 The first uncertainty is statistical and the second is systematic. Results from Belle are preliminary.}
 \label{tab:results}
\end{table}

\section{\boldmath Extraction of $\fds$ and Conclusions}
The value of $\fds$ is determined from measured branching fractions of leptonic $\ds$ decays by inverting Eq. \ref{eq:brleptonic_sm}.
The external inputs needed in the extraction of $\fds$ are all very precisely measured and do not introduce additional uncertainties 
except the $\ds$ lifetime, $\tau_{\ds}$, which introduces an 0.70\% relative uncertainty on $\fds$.

An error-weighted averages\footnote{Average of the decay constants extracted from measured $\br(\dsmunu)$ and $\br(\ds^+\to\tau^+\nu_{\tau})$.} of $\ds$-meson decay
constant, $\fds$, are found by Belle and BaBar to be
\begin{eqnarray}
 \fds^{\rm Belle} & = & (255.0\pm4.2(\rm stat.)\pm4.7(\rm syst.)\pm1.8(\tau_{D_s}))~\mbox{MeV}\\
 \fds^{\rm BaBar} & = & (258.6\pm6.4(\rm stat.)\pm7.3(\rm syst.)\pm1.8(\tau_{D_s}))~\mbox{MeV}.
\end{eqnarray}
Preliminary results from Belle represent the most precise measurement of $\fds$ up to date at single experiment. 

Averaging measurements of $\fds$ from $B$-factories with the one performed by CLEO-c experiment~\cite{Naik:2009tk}, 
$\fds^{\rm CLEO-c}  =  (259.0\pm6.2(\rm stat.)\pm2.4(\rm syst.)\pm1.8(\tau_{D_s}))~\mbox{MeV}$, gives an
experimental world average,
\begin{equation}
 \fds^{\rm WA}  =  (257.2\pm4.5)~\mbox{MeV},\\
\end{equation}
which is found to be within $2\sigma$ consistent with currently most precise lattice QCD calculation from HPQCD collaboration~\cite{Na:2012iu,Na:2013ti}, 
$\fds^{\rm HPQCD}=(246.0\pm3.6)$~MeV.

\end{document}